\documentclass{aa}  

\usepackage{hyperref}
\usepackage{graphicx}
\usepackage{mathrsfs}
\usepackage{txfonts}
\begin{document}

   \title{The Effect of Gravitational Stratification on Kink Oscillations in Curved Coronal Loops}

  \titlerunning{Kink Waves in Gravitational Coronal Loops}  
  
   \author{Mingzhe Guo\inst{1,2}
          \and
          Bo Li\inst{2}
          \and
          Mijie Shi\inst{2}
          }

   \institute{Institute of Frontier and Interdisciplinary Science, Shandong University, Qingdao, 266237, China\\
   \email{m.guo@email.sdu.edu.cn}
   \and
    Shandong  Key Laboratory of Space Environment and Exploration Technology, Institute of Space Sciences, Shandong University, Shandong, China
            }
   \date{Received; accepted}

 
  \abstract
   {Kink oscillation frequency is a key parameter for coronal seismology.
   It is still unclear how gravitational stratification affects the kink frequency in curved coronal loops.}
   {This work aims to investigate the effect of gravitational stratification on the frequency of kink oscillations in curved coronal loops and {
    discuss their seismological potential}.}
   {We conduct numerical computations within the ideal MHD framework to study {different kink polarizations and harmonics} in a curved, gravitationally stratified coronal loop. 
   The oscillation frequencies derived from the Lagrangian displacement are compared with the WKB approximation.}
   {For the {vertically polarized fundamental mode}, 
   the oscillation frequency deviates from the WKB approximation by about 18\% {in the current numerical setup}.
Nevertheless, the oscillation frequency closely matches the local Alfv\'en frequency near the loop apex. 
   {On the other hand,}
    the frequency of the horizontally polarized fundamental mode exhibits only a 7\% deviation {in our current model} from the WKB approximation
   and closely matches the local Alfv\'en frequency near one quarter of the loop.
   {For the first overtones,
   the frequencies for both polarizations can be well described by the WKB approximation.}}
   {The frequency of vertically polarized {fundamental kink modes} can be predicted by the local Alfv\'en frequency near the loop apex.
   In contrast, the WKB approximation remains highly reliable for estimating the frequency of horizontally polarized {fundamental modes and first overtones},
   which is also well described by the local Alfv\'en frequency near one quarter of the loop.
   These results therefore pave the way for spatially dependent coronal seismology, enabling, e.g., the probing of magnetic field strength at different locations along a coronal loop.}
   

   \maketitle
   
\nolinenumbers

\section{Introduction}

By combining
magnetohydrodynamic (MHD) wave theory with observations, 
coronal seismology provides inversion schemes to indirectly infer some physical parameters,
such as the coronal magnetic field \citep[see, e.g.,][for a recent review]{2020ARA&A..58..441N}.
As the only MHD wave mode that can disturb the axis of magnetic structures,
kink waves can be used to diagnose the coronal magnetic field based on their measured
frequencies regardless of standing or propagating waves \citep[e.g.,][]{2001A&A...372L..53N, 2020Sci...369..694Y, 2020ScChE..63.2357Y, 2024Sci...386...76Y}.
However,
current seismology schemes are still based on the theoretical framework of kink modes in canonical magnetic cylinders \citep[e.g.,][]{1975IGAFS..37....3Z,1983SoPh...88..179E}.
Gravity and curvature of coronal loops are essential, but have not yet been taken into account simultaneously in the context of coronal seismology. 

As a typical magnetic structure in the solar corona,
curved loops are known to be typical waveguides for kink waves \citep[e.g.,][]{2014LRSP...11....4R, 2020ARA&A..58..441N}.
The influence of loop curvature on kink modes was initially explored in two-dimensional (2D) models
\citep[e.g.,][]{1997A&A...317..752S,2005A&A...438..733B,2005A&A...440..385S,2006A&A...446.1139V,2006A&A...449..769V,2006A&A...452..615V}.
{These studies were }restricted to the examination of vertical polarizations,
{which can be excited, for example, by slow modes in short loops \citep[][]{2024MNRAS.527.5741L} or  vortex shedding 
\citep[][]{2021ApJ...908L...7K}}.
{Although vertical polarizations have also been confirmed in observations
\citep[e.g.,][]{2004A&A...421L..33W,2008A&A...489.1307W}, 
the other polarization, 
namely the more frequently observed horizontal kink polarization
\citep[e.g.,][]{2023NatCo..14.5298Z}, 
should also be examined. }
Analytically,
\citet{2004A&A...424.1065V,2009SSRv..149..299V} found that the curvature in a toroidal loop model has little influence on kink eigenmodes under a force-free magnetic field.
Numerically,
the frequency difference between horizontally and vertically polarized kink oscillations has been found not to be observable in a semi-torus model \citep{2006ApJ...650L..91T}.
Considering the WKB approximation,
\citet{2020ApJ...894L..23M} showed that the analytical prediction can reasonably describe the kink frequency in their planar model.
More recently,
\cite{2024A&A...687A..30G} examined the oscillation and damping properties for both horizontal and vertical polarizations, 
comparing their results with previous studies of different kink polarizations \citep{2009A&A...506..885R,2020ApJ...904..116G}.
However,
these studies on kink oscillations in curved loop models do not consider the effect of gravitational stratification.

In long coronal loops whose lengths significantly exceed the gravitational scale height, 
gravitational stratification is expected to influence loop dynamics. 
{Density stratification affects the displacements of fundamental kink modes in coronal loops \citep[e.g.,][]{2007A&A...462..743E}.}{
\citet{2005A&A...430.1109A} analytically investigated the influence of longitudinal density stratification on the frequencies and damping rates of kink eigenmodes.
Furthermore, 
based on deviations of the period ratio between the fundamental mode and the first overtone (i.e., $P_1/P_2$) from 2,
the density stratification can be determined \citep[][]{2005ApJ...624L..57A,2007A&A...473..959V}.
However, 
magnetic stratification induced loop expansion can increase $P_1/P_2$ and thus significantly affect the results of coronal seismology \citep[][]{2008A&A...486.1015V}.}
Regarding the propagation of kink modes, the effect of gravitational stratification was analytically investigated by \citet{1981A&A....98..155S}, who introduced a global cutoff. 
Although \citet{2014SoPh..289.3033L} offered a different perspective of the cutoff for kink waves in a stratified atmosphere, recent numerical work by \citet{2023A&A...672A.105P} has helped to clarify the physical scenario.
In recent numerical studies of kink oscillations in magnetic cylinders, gravity is routinely included in the contexts of both seismology \citep[e.g.,][]{2024A&A...689A.195G,2025RAA....25a5010G} and coronal heating \citep[e.g.,][]{2019A&A...623A..53K}. 
However, none of these studies of kink waves in stratified loops have simultaneously incorporated the effects of loop curvature.

The combined effects of gravity and loop curvature on kink oscillations have not yet been investigated in the same model. 
In the current work, 
we examine kink eigenmodes in a curved coronal loop with the presence of gravity.
This paper is organized as follows. 
Section \ref{sec_model} describes the equilibrium and the numerical setup.
The simulation results are presented and discussed in Section \ref{sec_results} and Section \ref{sec_discussion}, 
followed by a summary of the present work in Section \ref{sec_summary}. 
\section{Model Description}
\label{sec_model}

We considered a potential magnetic field similar to that in \cite{2024A&A...687A..30G}.
In Cartesian coordinate system $(x,y,z)$,
the initial magnetic field is given by 
\begin{eqnarray}
\begin{split}
B_x(x,z) = B_0\cos\left(\displaystyle\frac{x}{\Lambda_B} \right)\exp\left(-\displaystyle\frac{z}{\Lambda_B} \right),\\
B_z(x,z) = -B_0\sin\left(\displaystyle\frac{x}{\Lambda_B} \right)\exp\left(-\displaystyle\frac{z}{\Lambda_B} \right),
\end{split}
\label{eq_B}
\end{eqnarray}
where $B_0=30$G,
$\Lambda_B=2L_B/\pi$ with $L_B=145$Mm.
We also define a set of orthonormal basis vectors $(\hat{\vec{e}}_{\rm t}, \hat{\vec{e}}_{\rm h}, \hat{\vec{e}}_{\rm v})$,
\begin{equation}
\label{eq_basisVec}
\begin{split}
& \hat{\vec{e}}_{\rm t} = \cos(x/\Lambda_B)\hat{x}-\sin(x/\Lambda_B)\hat{z},  \\ 
& \hat{\vec{e}}_{\rm h} = \hat{y},                                            \\
& \hat{\vec{e}}_{\rm v} = \sin(x/\Lambda_B)\hat{x}+\cos(x/\Lambda_B)\hat{z}. 
\end{split}
\end{equation}
In the $x-z$ plane,
$\hat{\vec{e}}_{\rm t}$ ($\hat{\vec{e}}_{\rm v}$) is locally parallel (perpendicular) to the magnetic field.

We then consider a vertically stratified atmosphere.
The gravity is given by
\begin{eqnarray}
\vec{g}(z) = -\frac{g_{\odot}R^2_{\odot}}{(z+R_{\odot})^2}\hat{z},
\label{eq_grav}
\end{eqnarray}
where $g_{\odot}$ is the gravitational acceleration on the solar surface, 
$R_{\odot}$ is the solar radius.
The stratified background pressure and density are described by
\begin{eqnarray}
\frac{{\rm d}p(z)}{{\rm d}z} = -\frac{g_{\odot}R^2_{\odot}}{(z+R_{\odot})^2}\rho(z).
\label{eq_rho_prs}
\end{eqnarray}
{Note that the background temperature is constant.} 
Combined with the ideal gas law,
we can readily obtain the density and pressure of the background 
\begin{eqnarray}
\rho_{\rm e}(z) = \rho_{{\rm e}0} \exp\left(\frac{-g_{\odot}R_{\odot}}{A_0T_{\rm e}}+\frac{g_{\odot}R^2_{\odot}}{A_0T_{\rm e}}\frac{1}{z+R_{\odot}}\right),
\label{eq_rho_e}
\end{eqnarray}
\begin{eqnarray}
p_{\rm e}(z) = A_0T_{\rm e}\rho_{\rm e}(z),
\label{eq_prs_e}
\end{eqnarray}
where $A_0=2k_{\rm B}/m_{\rm H}$, $T_{\rm e}=2$MK, $\rho_{{\rm e}0}=4.18\times10^{-16}{\rm g ~cm^{-3}}$.

\begin{figure*}
   \centering
   \includegraphics[width=1.0\hsize]{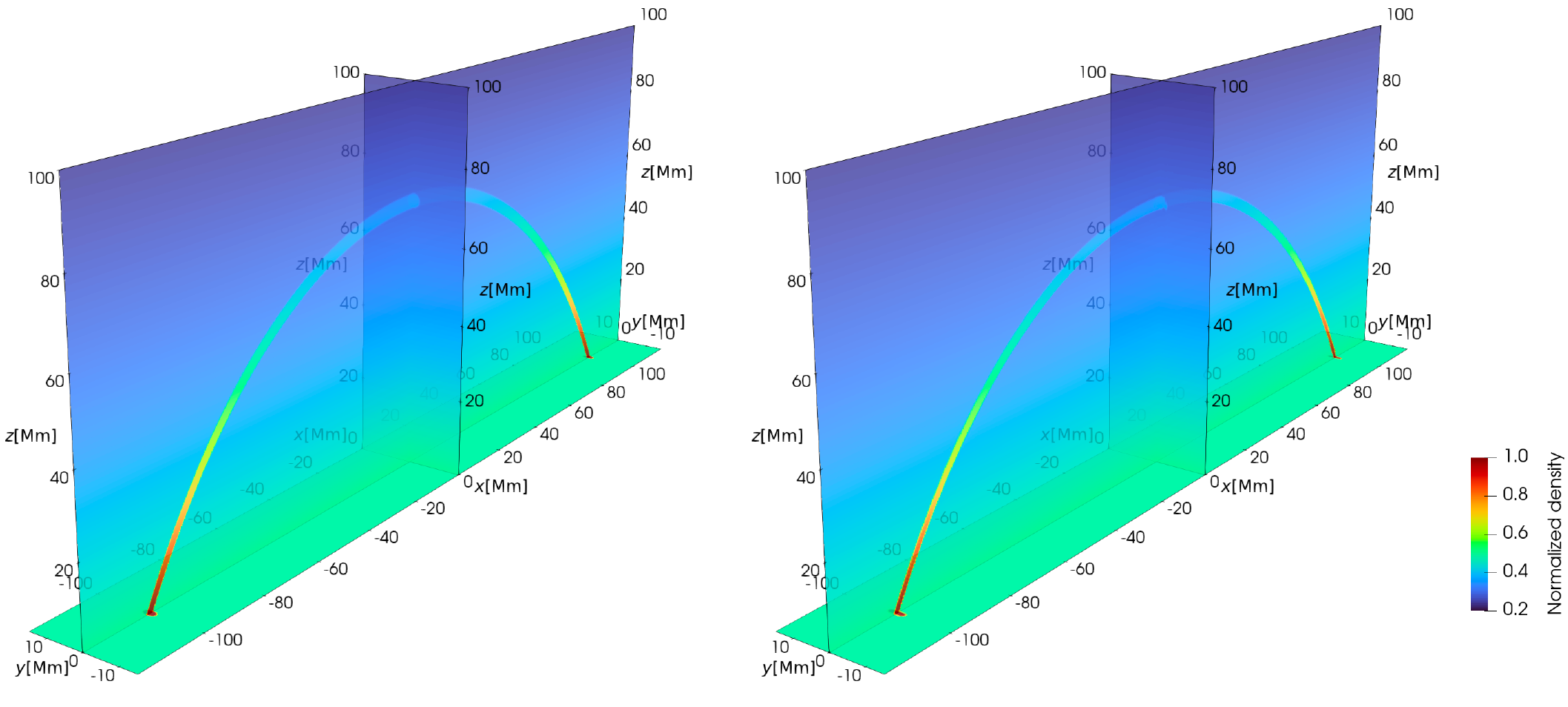}
   \caption{3D rendering of the density-enhanced loop. The left panel shows the initial state before relaxation, and the right panel shows the relaxed state.}
    \label{fig_loop_render}
    \end{figure*}

The density-enhanced loop is then constructed.
The internal loop density and pressure profiles at a given height $z$ are 
\begin{eqnarray}
  \rho_{\rm i}(z) = 2\rho_{{\rm e}0} \exp\left(\frac{-g_{\odot}R_{\odot}}{A_0T_{\rm i}}+\frac{g_{\odot}R^2_{\odot}}{A_0T_{\rm i}}\frac{1}{z+R_{\odot}}\right),
  \label{eq_rho_i}
\end{eqnarray}
\begin{eqnarray}
   p_{\rm i}(z)=A_0T_{\rm i}\rho_{\rm i}(z),
\end{eqnarray}
where $T_{\rm i}=1$MK.
This ensures a density ratio of 2 between the internal loop region and the external corona at $z=0$.
The density and pressure distributions are given by
\begin{eqnarray}
\rho(\Bar{r}) = \rho_
{\rm e}(z)+\displaystyle\frac{\rho_{\rm i}(z)-\rho_{\rm e}(z)}{2}\left\{1-\tanh\left[\left(\displaystyle\frac{\Bar{r}}{R}-1\right)b\right]\right\},
\label{eq_rho}
\end{eqnarray}
\begin{eqnarray}
p(\Bar{r}) = p_
{\rm e}(z)+\displaystyle\frac{p_{\rm i}(z)-p_{\rm e}(z)}{2}\left\{1-\tanh\left[\left(\displaystyle\frac{\Bar{r}}{R}-1\right)b\right]\right\},
\label{eq_prs}
\end{eqnarray}
where
\begin{eqnarray}
\Bar{r}= \sqrt{\left(\displaystyle\frac{\psi-\psi_0}{\Delta\psi}\right)^2+\left(\displaystyle\frac{y}{\Delta y}\right)^2},
\label{eq_r}
\end{eqnarray}
with 
\begin{eqnarray}
\psi=B_0\Lambda_B\cos\left(\displaystyle\frac{x}{\Lambda_B}\right)\exp\left(-\displaystyle\frac{z}{\Lambda_B}\right),
\label{eq_psi}
\end{eqnarray}
and $\psi_0=\psi(x=100{\rm Mm},z=0)$,
$\Delta\psi=B_0\Lambda_B$,
$\Delta y=2\Lambda_B$.
Here, $R={\cos\left(100/\Lambda_B\right)-\cos\left(101/\Lambda_B\right)}$ prescribes
   the $x$-extent of the loop at its footpoints. 
The parameter $b=10$ determines the width of the boundary layer.
Note that the density enhanced loop is in a low-$\beta$ environment with a maximum plasma $\beta$ of 0.016.
The 3D view of the initial state is shown in
Figure~\ref{fig_loop_render}a.


Before proceeding,
we need to specify our numerical setup.
Regarding the bottom boundary at $z=0$,
all the components of the velocity are set to vanish,
while all the components of the magnetic field follow the zero gradient condition.
The density and pressure are fixed at their initial values.
{Outflow boundary conditions are applied to all the other side boundaries.}
To ensure the stability of the computation,
we also consider a velocity absorption layer \citep[see also][]{2023A&A...672A.105P,2023ApJ...949L...1G,2024A&A...689A.195G}.
The velocity is modified by  
\begin{eqnarray}
v'(x,y,z;t)=\alpha(t)\gamma(z) v(x,y,z;t),
\label{eq_vnew}
\end{eqnarray}
where 
\begin{eqnarray}
  \alpha (t)
= \left\{
       \begin{array}{ll}
          0.998+0.002\displaystyle\frac{t}{t_c},   & 0 < t \le t_{\rm c},  \\
           1,   & t > t_{\rm c},
        \end{array} 
  \right.
\label{eq_alpha}
\end{eqnarray}
and 
\begin{eqnarray}
\gamma(z)= 0.5\left\{1-\tanh\left[0.3\left(\frac{z-90}{l_0}\right)\right]\right\},
\label{eq_beta}
\end{eqnarray}
with $t_c=2000$s and $l_0=1$Mm being the unit of length.
We then solve the set of ideal 3D MHD equations with the PLUTO code \citep{2007ApJS..170..228M}.
As in \cite{2024A&A...687A..30G},
a similar magnetic field splitting method is used.
We choose the piecewise parabolic scheme for spatial reconstruction,
the HLLD Riemann solver to calculate the numerical fluxes, 
and the second-order Runge–Kutta method for time marching.
{The computational domain is $[-120,120]\times[-15,15]\times[0,100]$Mm.
We adopt a set of uniform grid with $600\times120\times400$ cells.}

{Note that relaxation is necessary, as it is not a magnetostatic balance system so far.}
The system reaches equilibrium after a relaxation period of 2000s, 
and the relaxed state is shown in Figure~\ref{fig_loop_render}b.
In the following, we consider this relaxed state as $t=0$.
The maximum residual velocity in the system is less than {$2{\rm km/s}$.}
Since the characteristic evolution timescale of the system,
$\lvert {\delta \rho /\left[\nabla\cdot(\rho\vec{v})\right]} \lvert$,
is much longer than the period of the wave of interest, 
the system can thus be regarded as a stable state.

Then we specify the initial perturbations.{
We distinguish two different polarizations, 
namely horizontal and vertical.
As discussed in \cite{2024A&A...687A..30G},
the density distribution across the loop cross-section is no longer axisymmetric due to gravity induced stratification.
Since the density remains nearly constant in the horizontal direction at a given height,
horizontally polarized modes experience little influence from stratification.
In contrast, 
vertically polarized kink modes are expected to be more strongly affected by gravitational stratification.
}
In the following, we first focus on vertical polarization.
Similar to the initial conditions in \citet{2024A&A...687A..30G},
we introduce an initial velocity perturbation of the form
\begin{eqnarray}
  \vec{v}(x,y,z; t=0) 
= \displaystyle\frac{v_0\sin{\theta}}{2}\left\{1-\tanh\left[\left(\displaystyle\frac{\Bar{r}}{R}-1\right)b\right]\right\} \hat{\vec{e}}_{\rm v},
  \label{eq_vv}
\end{eqnarray}
where $v_0=20{\rm km/s}$ is the amplitude of the velocity perturbation,
$\sin\theta=\dfrac{z}{\sqrt{x^2+z^2}}$ with $\theta$ being the angle along the loop from the bottom.
It ensures maximum velocity perturbation at loop apex and zero velocity at loop footpoints. 
For reference,
we also consider a horizontal velocity perturbation in the 
$y$-direction
\begin{eqnarray}
  \vec{v}(x,y,z; t=0) 
= \displaystyle\frac{v_0\sin{\theta}}{2}\left\{1-\tanh\left[\left(\displaystyle\frac{\Bar{r}}{R}-1\right)b\right]\right\} \hat{\vec{e}}_{\rm h}.
  \label{eq_vh}
\end{eqnarray}
{In addition,
we also consider another set of initial velocity conditions to excite the first overtones.
The velocity perturbation in the vertical direction is given by
\begin{eqnarray}
  \vec{v}(x,y,z; t=0) 
= v_0\sin{\theta}\cos{\theta}\left\{1-\tanh\left[\left(\displaystyle\frac{\Bar{r}}{R}-1\right)b\right]\right\} \hat{\vec{e}}_{\rm v},
  \label{eq_vv1}
\end{eqnarray}
and the horizontal velocity perturbation is given by 
\begin{eqnarray}
  \vec{v}(x,y,z; t=0) 
= v_0\sin{\theta}\cos{\theta}\left\{1-\tanh\left[\left(\displaystyle\frac{\Bar{r}}{R}-1\right)b\right]\right\} \hat{\vec{e}}_{\rm h}.
  \label{eq_vh1}
\end{eqnarray}
}
\section{Results}
\label{sec_results}
   \begin{figure}
   \centering
   \includegraphics[width=1.0\hsize]{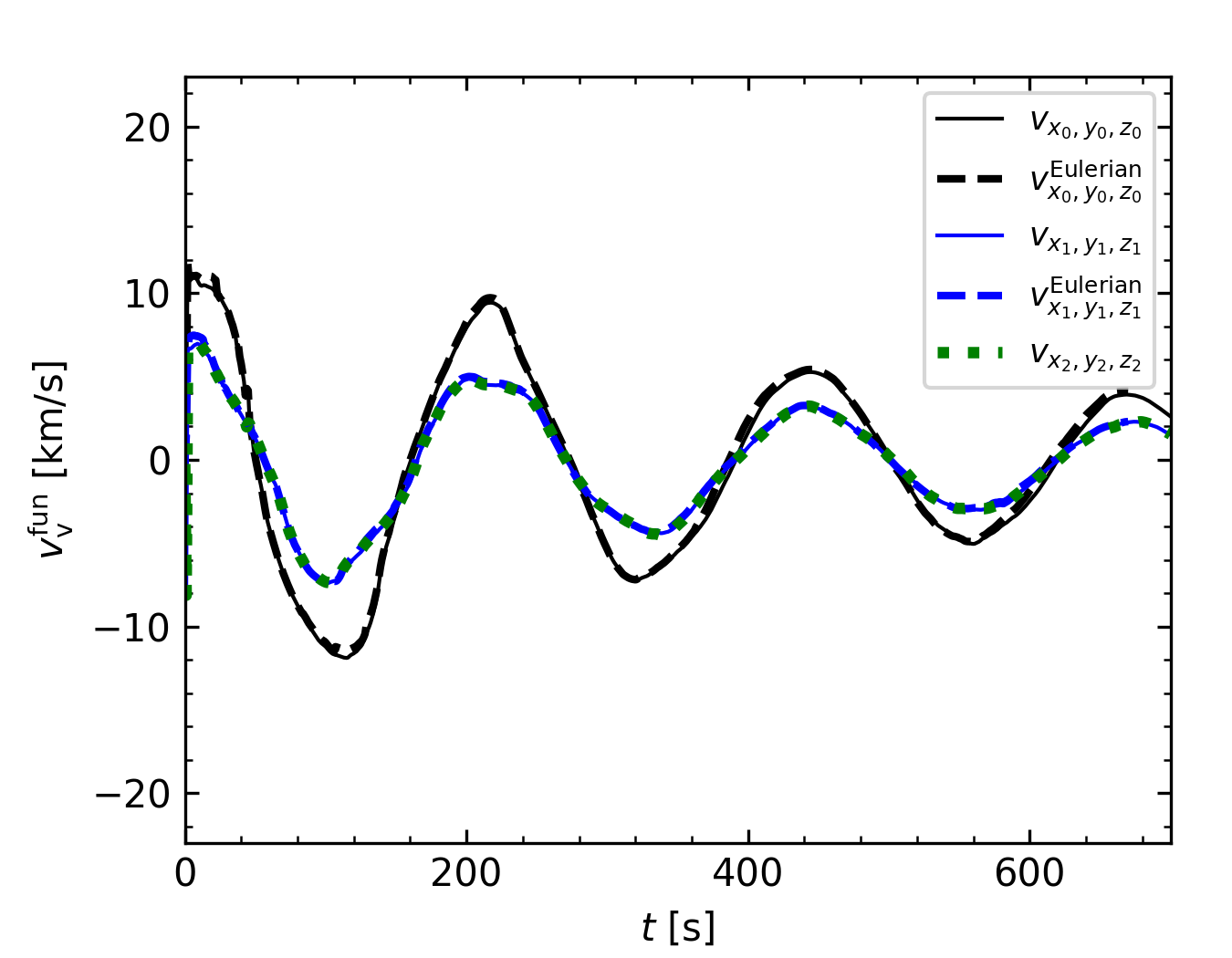}
   \caption{{
Temporal evolution of $v_{\rm v}$ for the vertically polarized fundamental kink oscillations.
The black, blue, and green lines represent $v_{\rm v}$ of fluid parcels initially located at the loop apex ($[x_0,y_0,z_0]=[0,0,70]$~Mm), at one quarter of the loop ($[x_1,y_1,z_1]=[59.4,0,49.4]$~Mm), and at the opposite quarter of the loop ($[x_2,y_2,z_2]=[-59.4,0,49.4]$~Mm), respectively.
The black and blue dashed line designate $v_{\rm v}$ measured at the fixed position $[x_0,y_0,z_0]$ and $[x_1,y_1,z_1]$.}}
    \label{fig_v_v}
    \end{figure}  
   \begin{figure}
   \centering
   \includegraphics[width=1.0\hsize]{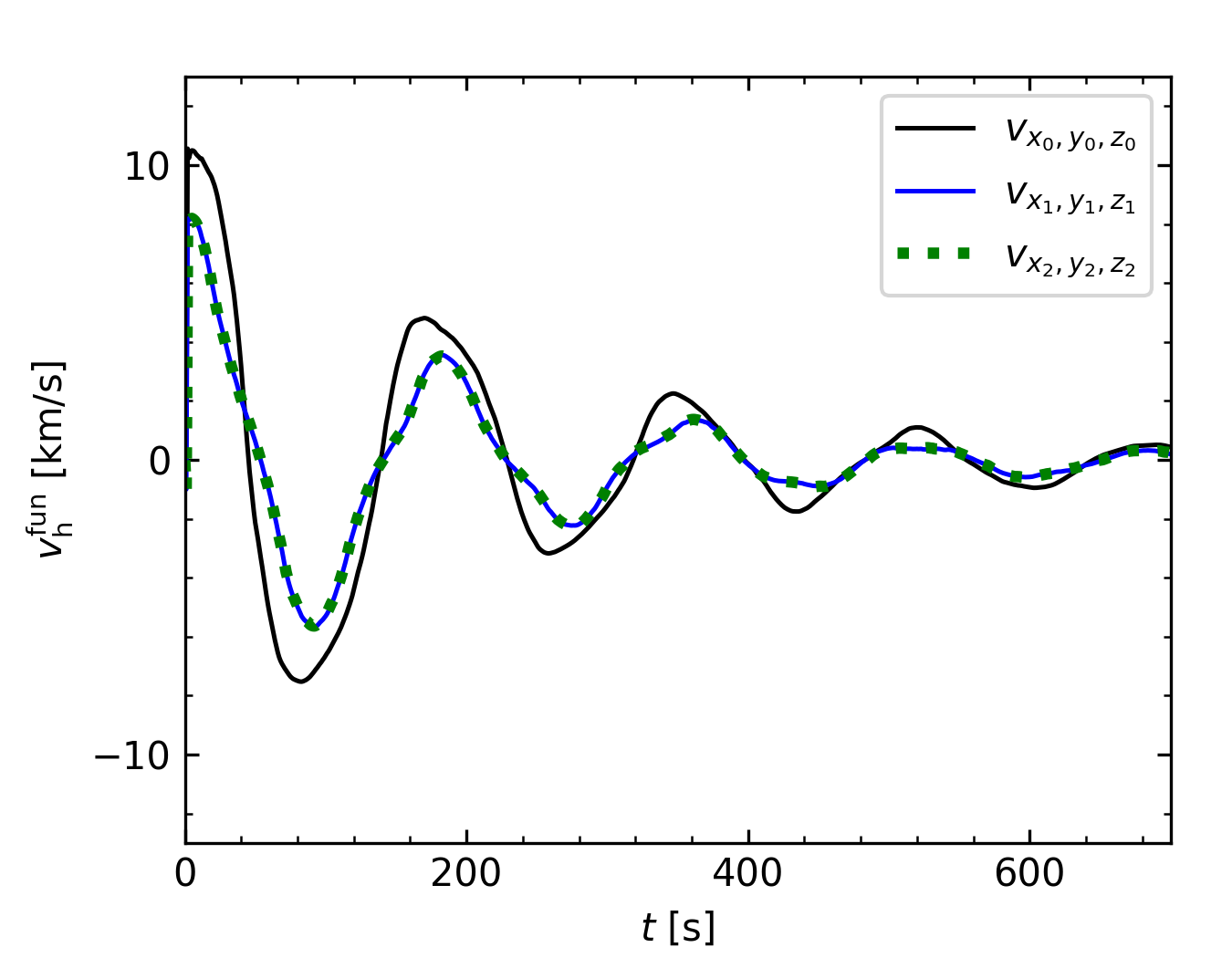}
   \caption{Similar to Figure~\ref{fig_v_v}, but for the horizontal polarized fundamental kink oscillations.}
    \label{fig_v_h}
    \end{figure}  

We first examine the frequency of the vertical polarization.
To better compare the results with analytical predictions,
we track the Lagrangian displacement of a fluid parcel.
This is implemented by assigning passive {tracer} in the code.
As such,
we can follow the displacement of a fluid parcel within the loop region.
The solid lines in Figure~\ref{fig_v_v} show the temporal evolution of the Lagrangian velocity of the specified fluid parcels.
Different fluid parcels in the loop are sampled,
as distinguished by different colors.
One fluid parcel is initially located at the loop apex ($[x, y, z] = [0, 0, 70]\ \mathrm{Mm}$) at $t = 0$,
{the other two fluid parcels are initially located at one quarter position of the loop ($[x, y, z] = [59.4, 0, 49.4] \mathrm{Mm}$) and the opposite quarter of the loop ($[x, y, z] = [-59.4, 0, 49.4] \mathrm{Mm}$)}.
Following a procedure similar to that in \citet{2020ApJ...904..116G, 2024A&A...687A..30G},
we estimate the oscillation period as twice the average time interval between two adjacent extrema of the oscillating profile.
For the loop apex (black {solid} curve),
this procedure yields a period of {$P^{\rm fun}_{\rm v}=221.6$}s,
while the oscillation period {at the other two quarters} of the loop is {226}s.
{Neglecting the small measurement errors at different locations, 
we find that the periods are similar. }
{We also find that these oscillation profiles are in phase,
and that the two quarters of the loop exhibit the same oscillation behaviour}.
This clearly indicates that the excited oscillation in the current loop corresponds to the fundamental mode.
To be consistent with observations,
we use the period at the loop apex to represent the oscillation period in the loop hereafter.

Under some assumptions in a straight magnetic cylinder, 
such as the thin-tube-thin-boundary (TTTB) approximation,
the period can be predicted through modal analysis \citep[e.g.,][]{1983SoPh...88..179E,1992SoPh..138..233G}.
When the Alfv\'en frequency varies along the loop axis,
the oscillation frequency can be estimated using the WKB approximation.
For loops with an axisymmetric density profile across the loop, 
the WKB approximation can describe the oscillation frequency quite well
\citep[][]{2024A&A...687A..30G}.
In the current model, 
applying the WKB approximation yields a predicted period
\begin{eqnarray}
P_{\rm WKB}=2\int^L_0 \frac{{\rm d}s}{c_{\rm k}(s)}=2\int^L_0\sqrt{\frac{\rho_{\rm i}(s)+\rho_{\rm e}(s)}{2\rho_{\rm i}(s)}} \displaystyle\frac{{\rm d}s}{v_{\rm Ai}\left(s\right)},
\label{eq_WKB}
\end{eqnarray}
with 
\begin{eqnarray}
v_{\rm Ai}\left(s\right)=\displaystyle\frac{B\left(s\right)}{\sqrt{\mu_0 \rho_{\rm i}\left(s\right)}},
\label{eq_WKB_alfven}
\end{eqnarray}
where $c_{\rm k}$ is the kink speed, 
$v_{\rm Ai}$ is the internal Alfv\'en speed along the loop axis,
$L$ represents the loop length, 
$s$ is the distance from a footpoint along the loop axis and can be expressed as a function of $z$,
and $B(s)$ is the magnetic field strength along $s$.
According to this formula,
we thus obtain a period of {$P_{\rm WKB}=188.3$} s.
Therefore, the oscillation period $P_{\rm v}$ deviates from this theoretical estimation ($P_{\rm WKB}$) by approximately $18\%$.

In general, 
the Lagrangian velocity is expected to be different from the Eulerian velocity measured at a fixed position. 
However,
in the current case, 
Figure \ref{fig_v_v} shows that the Lagrangian velocity closely matches the velocity measured at the fixed positions (filled black circles in Figure~\ref{fig_v_v}). 
This agreement can be understood by calculating the Lagrangian displacement from the perturbation amplitude and oscillation period. 
The resulting displacement is still smaller than the cross-sectional radius at the loop apex, 
confirming that the traced fluid parcel remains within the internal loop region with uniform velocity.
In the following analysis,
we focus on the Lagrangian velocity only.

Figure~\ref{fig_v_h} shows horizontal oscillation profiles similar to those in Figure~\ref{fig_v_v}.
We can also compute the oscillation period from the different positions of the loop.
{At the apex (one quarter) of the loop,
the period is measured as {$P^{\rm fun}_{\rm h}=174.8$} s (178 s).
The similar periods,
as well as the in phase property of different positions again indicates that the excited horizontal oscillation in the current loop is an axial fundamental mode.
We use the period at the loop apex to represent the oscillation period.}
Therefore,
the oscillation period of the horizontal polarization deviates from $P_{\rm WKB}$ 
by only {$7\%$}.

   \begin{figure}
   \centering
   \includegraphics[width=1.0\hsize]{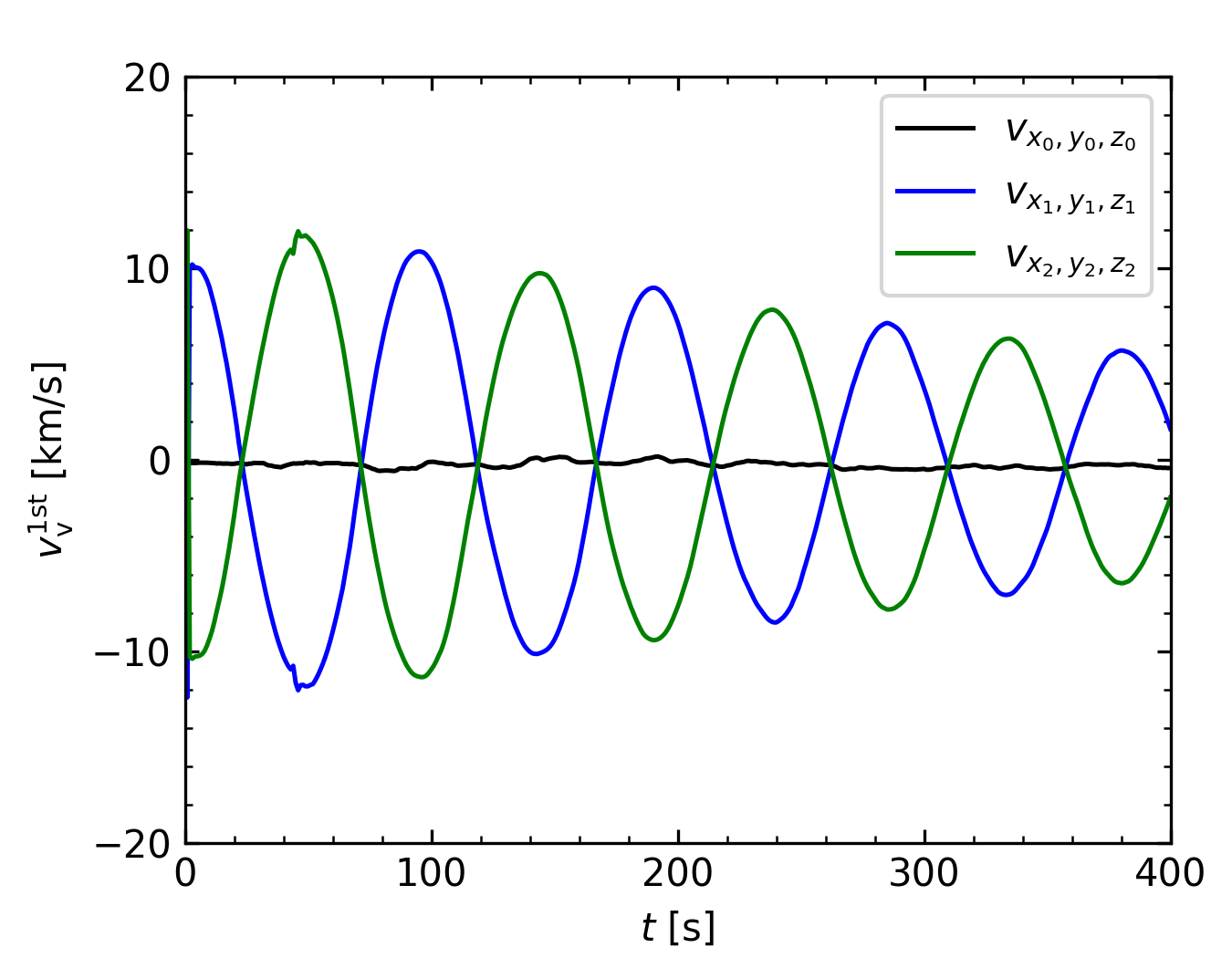}
   \caption{Temporal evolution of $v_{\rm v}$ for the vertically polarized first overtones of kink oscillations.
The black, blue, and green lines represent $v_{\rm v}$ of fluid parcels initially located at the loop apex ($[x_0,y_0,z_0]=[0,0,70]$~Mm), at one quarter of the loop ($[x_1,y_1,z_1]=[59.4,0,49.4]$~Mm), and at the opposite quarter of the loop ($[x_2,y_2,z_2]=[-59.4,0,49.4]$~Mm), respectively.}
   \label{fig_v_v_1st}
    \end{figure}  

   \begin{figure}
   \centering
   \includegraphics[width=1.0\hsize]{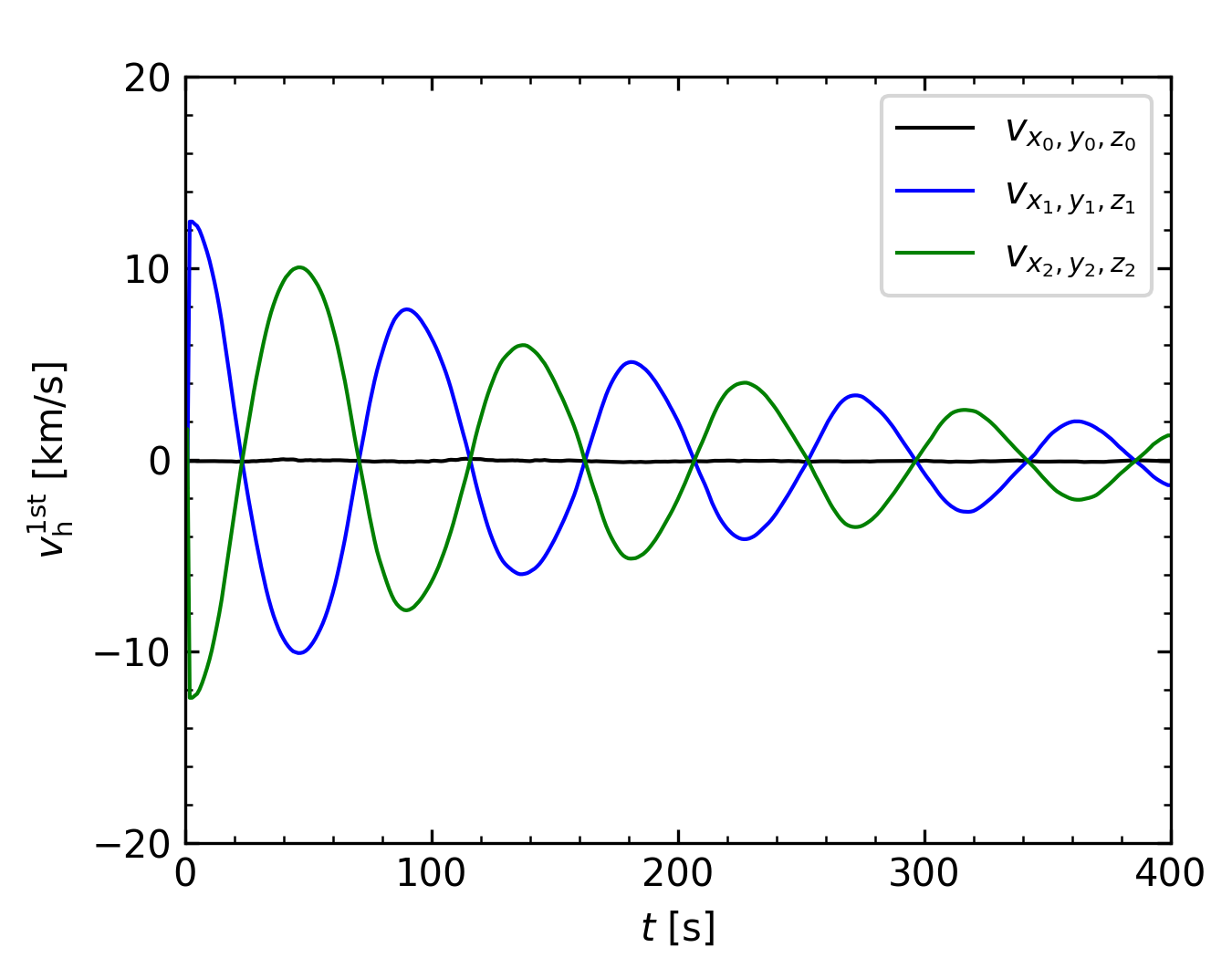}
   \caption{Similar to Figure~\ref{fig_v_v_1st}, but for the horizontal first overtones of kink oscillations.}
   \label{fig_v_h_1st}
    \end{figure}  

{
We also examine the oscillation profiles of the first overtones.
As for the fundamental modes, we consider both vertical and horizontal polarizations of the first overtones.
Oscillation profiles at the loop apex, as well as at one quarter and the opposite quarter of the loop are presented in Figures~\ref{fig_v_v_1st} and \ref{fig_v_h_1st}.
As expected for the first overtones,
the oscillation at one quarter of the loop is in antiphase with that at the opposite quarter.
The velocity at the loop apex is nearly zero, indicating that this location corresponds to a wave node.
Before proceeding, 
we estimate the predicted period of the first overtones,
which is half that of the fundamental modes, namely 94.15~s.
To determine the oscillation periods from the numerical results,
the same procedure described in Figure~\ref{fig_v_v} is applied to the oscillation profiles here.
We find that the oscillation period for the vertically polarized mode is {$P^{\rm 1st}_{\rm v}=95.4$}~s,
while that for the horizontally polarized mode is {$P^{\rm 1st}_{\rm h}=90.3$}~s.
Both values closely match the predicted period, 
with a maximum deviation of about $4\%$.
}

The damping of the oscillation profiles can also be observed.
It is widely accepted that resonant absorption is the dominant damping mechanism of kink oscillations \citep[e.g.,][]{2002A&A...394L..39G,2024A&A...687A..30G}.
However, 
resolving the resonant process within the boundary layer surrounding the loop requires a higher numerical resolution.
Therefore,
the damping process presented here may not be reliable.
Nevertheless, 
in the present work, 
we mainly focus on the periods of different polarizations and harmonics, 
which are not significantly influenced by numerical resolutions.

In addition, 
we note that the oscillation profiles in Figure~\ref{fig_v_h} and Figure~\ref{fig_v_h_1st} are not as sinusoidal as those in the other figures, showing some reverse steepening. 
This may result from the excitation of higher harmonics of the kink oscillations. 
However, a detailed examination is beyond the scope of the current work.

   \begin{figure*}
   \centering
   \includegraphics[width=1.0\hsize]{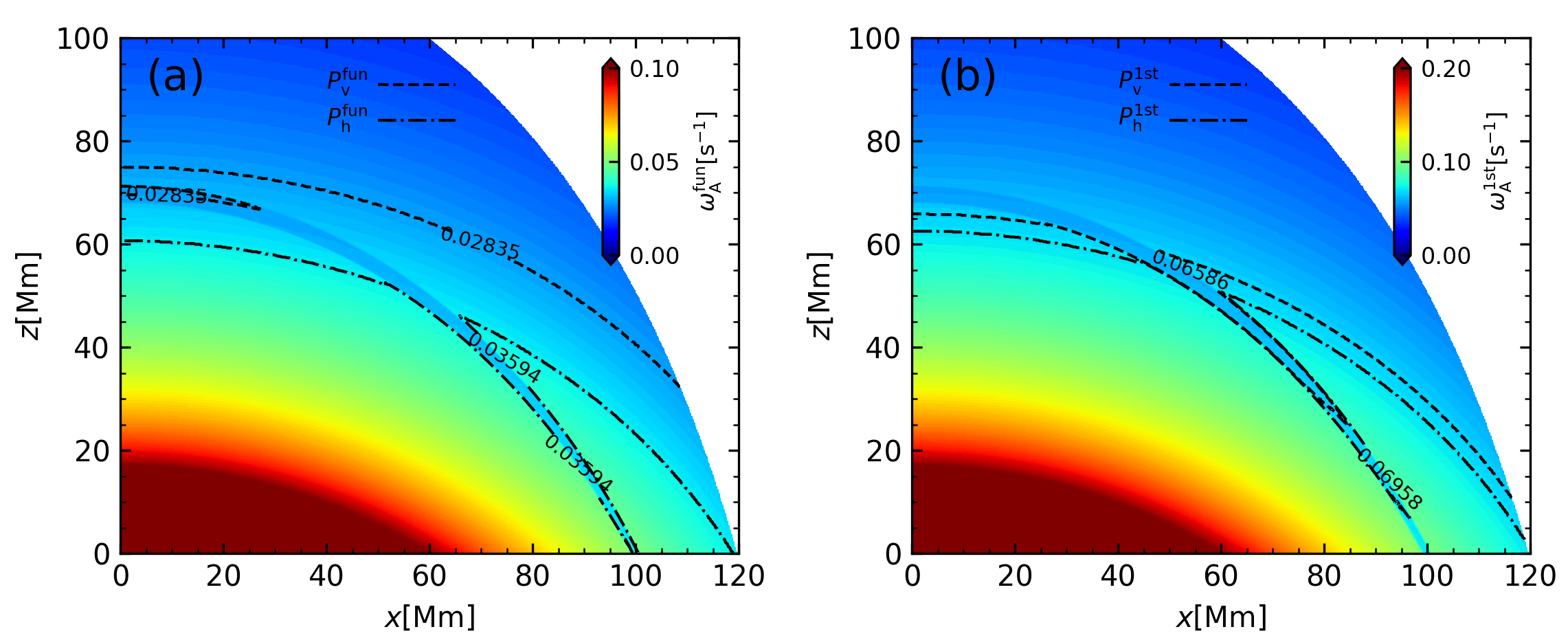}
   \caption{{
   Map of the local Alfv\'en frequency on the $y=0$ plane for (a) the fundamental modes and (b) the first overtones.
The dashed (dash-dotted) lines represent contours of the frequencies corresponding to the vertical (horizontal) polarizations as labelled.
} }
    \label{fig_alf_freq}
    \end{figure*}  

We further examine the distribution of the local Alfv\'en frequency to gain insight into the oscillation period.
Having the Alfv\'en speed distribution at hand,
we can calculate the local Alfv\'en frequency distribution $\omega_{\rm A}(x,z)$.
{For axial fundamental modes and the first overtones,
we have
\begin{eqnarray}
\omega^{\rm fun}_{\rm A}(x,z) = \frac{\pi}{\mathscr{L}(x,z)}v_{\rm A}(x,z),
\label{eq_alf_req}
\end{eqnarray}
and 
\begin{eqnarray}
\omega^{\rm 1st}_{\rm A}(x,z) = \frac{2\pi}{\mathscr{L}(x,z)}v_{\rm A}(x,z),
\label{eq_alf_req_1st}
\end{eqnarray}}
where $\mathscr{L}$ is the length of the field line anchored at bottom boundary, threading a given point $(x,z)$.
Figure~\ref{fig_alf_freq} shows the local Alfv\'en frequency distribution on the surface $y=0$.
{We also translate the periods of different kink polarizations to frequencies for both the fundamental modes and the first overtones.
In Figure~\ref{fig_alf_freq} (a),
the local Alfv\'en frequency distribution for the fundamental modes is illustrated.
The frequency of the vertically polarized kink mode matches the local Alfv\'en frequency near the loop apex,
as shown by the dashed black line.}
This suggests that 
we can use the local Alfv\'en frequency near the apex to predict the real oscillation frequency of the {vertically polarized fundamental modes}.
When we move downward along the loop,
the local Alfv\'en frequency increases.
Near one quarter of the loop, 
the local frequency becomes comparable to both the value calculated from the WKB approximation $P_{\rm WKB}$ and the measured period $P^{\rm fun}_{\rm h}$ of the horizontally polarized fundamental mode (dash-dotted line).
As aforementioned,
WKB theory can describe the frequency of horizontal modes quite well, 
as previously confirmed by \citet[][]{2024A&A...687A..30G}.
{Figure~\ref{fig_alf_freq} (b) shows the local Alfv\'en frequency distribution for the first overtones.
We also overplot the frequencies of both the vertically and horizontally polarized first overtones.
Since the periods of the two polarizations are similar and closely match the WKB prediction,
the frequencies for both polarizations can also be described by local Alfv\'en frequency near one quarter of the loop.}

Furthermore,
we can also see the dashed lines outside the loop region in Figure~\ref{fig_alf_freq}.
This leads to the wave tunneling effect and results in lateral leakage of the wave.
This scenario has been discussed in \cite{2024A&A...687A..30G}.

\section{Discussion}
\label{sec_discussion}

\subsection{{Frequency deviations from WKB theory}}
\label{sec_freq_WKB}

To understand the deviation of the oscillation frequency from the WKB approximation, 
we should revisit the local Alfv\'en frequency. 
For kink modes,
the oscillation frequency is determined by contributions from both the internal and external loop regions.
As shown in Figure \ref{fig_alf_freq},
the local Alfv\'en frequency decreases along the $\hat{e}_{\rm v}$ direction 
(perpendicular to the magnetic field lines). 
Therefore, for the vertically polarized kink mode, 
regions with both higher and lower frequencies than that inside the loop become relevant. 
{For vertically polarized fundamental modes,}
the damping of the oscillation (see Figure~\ref{fig_v_v}) leads to an unequal contribution from the lower and upper loop regions. 
Given that the realistic oscillation is typically excited below the loop \citep[e.g.,][]{2015A&A...577A...4Z}, 
leading to an initial upward motion, 
the upper loop region contributes more fractions. 
As a result, 
the measured oscillation frequency is lower than the WKB prediction.
This is also consistent with previous results in \cite{2024A&A...687A..30G} when a uniform density loop is considered.
In the current case,
the deviation from the WKB approximation is larger.
This is because gravitational stratification enhances the non-uniformity of the loop along its axis.
Consequently, a steeper gradient develops in the local Alfv\'en frequency,
which introduces greater uncertainty into the WKB approximation.
{For vertically polarized first overtones,
the antiphase oscillation with respect to the apex decreases the above effect.
Therefore,
the deviations of the vertically polarized first overtones from the WKB approximation are quite small.}
In the horizontal case, 
the loop oscillates in the $\hat{e}_{\rm h}$ direction of almost uniform local Alfv\'en frequencies, 
allowing the WKB approximation to predict the oscillation frequency with a smaller deviation.

{
The frequency of the vertically polarized fundamental mode matches 
the local Alfv\'en frequency near the loop apex.
In principle, 
the frequency calculated from the WKB approximation should correspond to the local Alfv\'en frequency at some location in the loop region, 
as indicated by Eq.~\eqref{eq_WKB} and illustrated in Figure~\ref{fig_alf_freq}.
In the current model, 
as discussed above, 
the measured frequency is lower than the WKB prediction, 
meaning that the oscillation frequency is shifted toward the loop apex.
This is not coincidental, 
but rather a natural consequence of vertically polarized fundamental modes in the current model.
}

\subsection{Implications for magnetic field measurements}
\label{sec_B}

The oscillation frequency serves as a key observable of kink modes in coronal loops.
However, 
in seismology processes in stratified coronal loops,
it should be emphasized 
that even when using a WKB approximation that incorporates axial non-uniformity (as in Section \ref{sec_results}), 
the inversion process inherently retains a degree of uncertainty. 
In the case of horizontally polarized kink modes,
this uncertainty remains acceptable, 
given that the observed oscillation frequency deviates from the WKB prediction by only about $5\%$. 
However,
seismological application for vertically polarized kink modes requires greater caution when based solely on the WKB theory. 
In realistic gravitationally stratified loops, 
axial non-uniformity may lead to significant deviations in oscillation frequency. 
Nevertheless,
we suggest that one should consider the local Alfv\'en frequency near the loop apex in potential seismology applications when the vertical kink polarization is involved to avoid large uncertainties.

{
The present results pave the way for spatially dependent inference of the coronal magnetic field strength.
In coronal seismology, 
the magnetic field strength can be inferred from the kink speed, provided that the phase speed and density are known \citep[e.g.,][]{2020Sci...369..694Y, 2020ScChE..63.2357Y, 2024Sci...386...76Y}.
In the current scenario, 
as discussed above, 
the measured periods of different fundamental polarizations reveal different local Alfv\'en frequencies.
This provides a potential approach to probe the local magnetic field strength once the local density is determined.
During the TRACE era, 
\cite{2008A&A...489.1307W} reported that different kink polarizations could coexist within the same coronal loop.
However, 
unambiguous observation of kink polarizations has only become possible in recent years with multi-angle and high-resolution observations \citep[e.g.,][]{2023NatCo..14.5298Z}.
In addition, 
higher temporal resolution observations may further help to distinguish the different periods associated with different polarizations.
}

\subsection{Scale height diagnostics}
\label{sec_p1p2}

   \begin{figure}
   \centering
   \includegraphics[width=1.0\hsize]{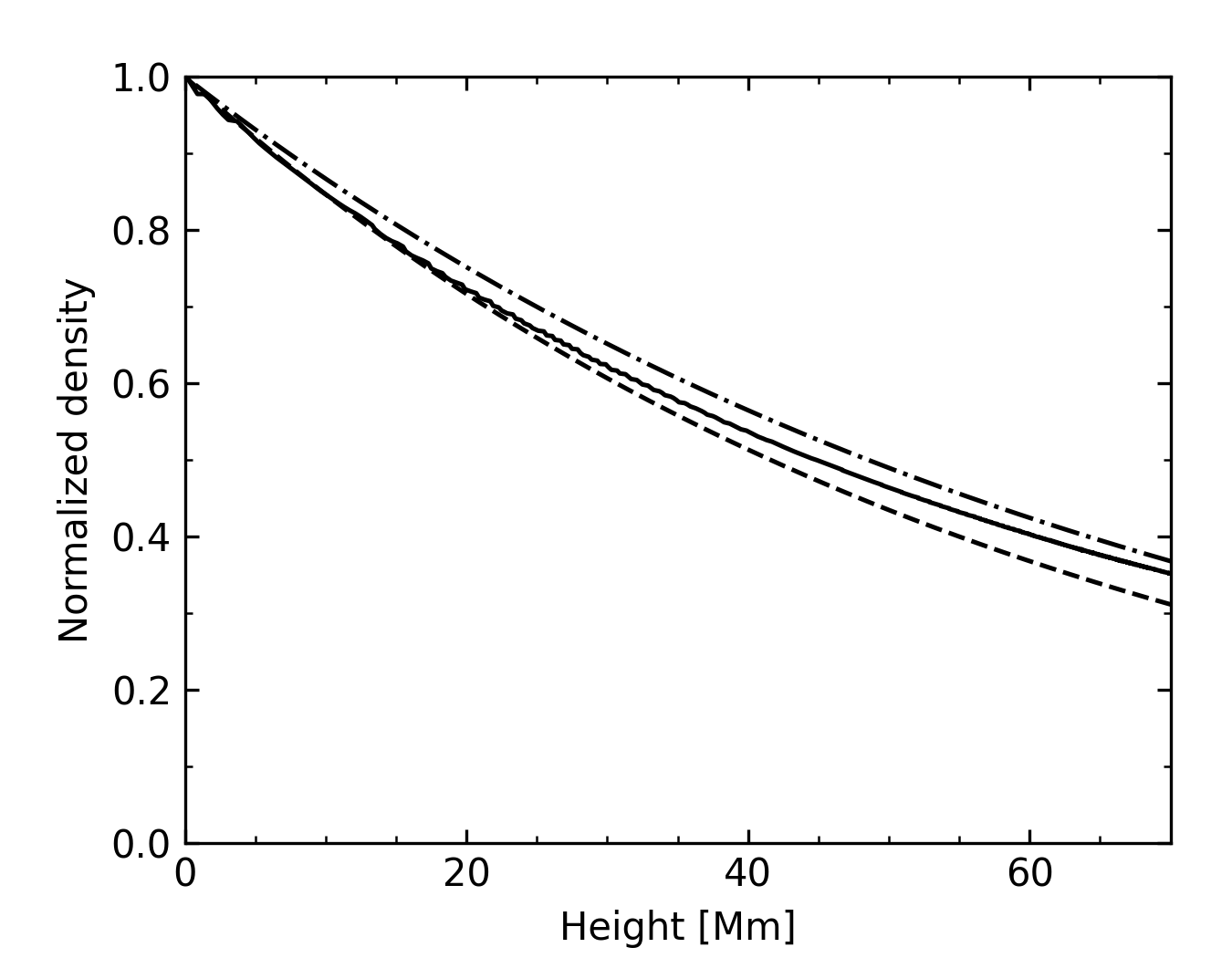}
   \caption{{Initial density distribution along the loop axis. Dashed and dash-doted lines illustrate the function of $\exp(-z/60)$ and $\exp(-z/70)$, respectively.}}
    \label{fig_ini_density}
    \end{figure}  

   \begin{figure}
   \centering
   \includegraphics[width=1.0\hsize]{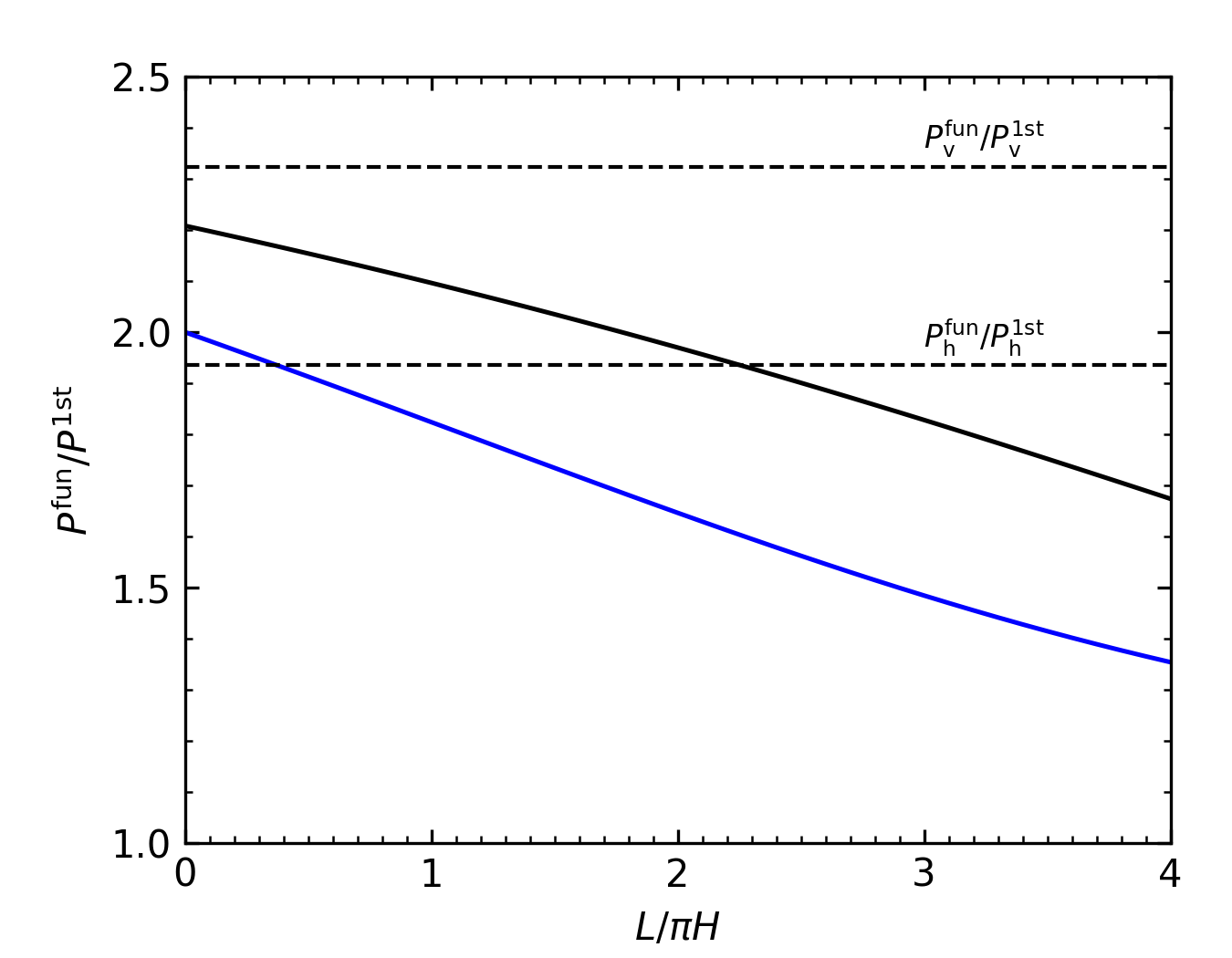}
   \caption{{The period ratio $P^{\rm fun}/P^{\rm 1st}$ as a function of the density stratification $L/\pi H$. The solid curves are derived from Equation~\eqref{eq_SL}. The dashed lines indicate the period ratio of $P^{\rm fun}_{\rm h}/P^{\rm 1st}_{\rm h}$ and $P^{\rm fun}_{\rm v}/P^{\rm 1st}_{\rm v}$ as labelled. See text for more details.}}
    \label{fig_p1p2}
    \end{figure}  

{It has been known that deviations of the period ratio $P_1/P_2$ (i.e., $P^{\rm fun}/P^{\rm 1st}$ in the present context) from 2 can be used to diagnose the density scale height in a stratified atmosphere \citep[e.g.,][]{2005ApJ...624L..57A,2007A&A...473..959V}. 
In our computations, 
in principle, 
the density scale height can be inferred since we have obtained the period ratio $P^{\rm fun}/P^{\rm 1st}$ for different polarizations.
Before proceeding,
we first examine the initial density distribution along the loop axis.
As shown in Figure~\ref{fig_ini_density}, 
the density profile does not follow an exact exponential form, 
but still lies between two exponential functions, 
namely $\exp(-z/60)$ and $\exp(-z/70)$.
This indicates that the density distribution is close to exponential, 
with a scale height between 60~Mm and 70~Mm.
According to \citet{2005ApJ...624L..57A},
the relationship between the period ratio $P^{\rm fun}/P^{\rm 1st}$ and the scale height $H$ can be derived.

From a different perspective,
inspired by \citet{2006A&A...457.1059D},
we consider solving the Sturm-Liouville problem 
\begin{eqnarray}
-\frac{{\rm d}^2 u(s)}{{\rm d} z^2}=\frac{\omega^2}{c^2_{\rm k}(s)}u(s), u(s)=0~{\rm at}~s=0,L.
\label{eq_SL}
\end{eqnarray}
The procedure is to provide the nonuniform distribution of $c_{\rm k}(s)$ and then find the eigenvalues $\omega^2$.
In \citet{2005ApJ...624L..57A},
a uniform magnetic field is assumed,
so the kink speed $c_{\rm k}$ is 
\begin{eqnarray}
c_{\rm k}(s)=\sqrt{\frac{2B^2_0}{\mu_0\rho(s)(1+\rho_{\rm ie})}},
\label{eq_ck_A05}
\end{eqnarray}
where $\rho_{\rm ie}=2$ is the density ratio,
$\rho(s) =\rho_0\exp[-L/{\pi H}\sin(\pi s/L)]$.
Then the period ratio $P^{\rm fun}/P^{\rm 1st}$ as a function of $L/\pi H$ presented in \citet{2005ApJ...624L..57A} can be reproduced, 
as shown by the blue line in Figure~\ref{fig_p1p2}.
However,
the magnetic field inhomogeneity in our current model has not been included.
We then take the variation of the magnetic field strength,
namely $B_0\exp\left[-L/{\pi \Lambda_B}\sin(\pi s/L)\right]$,
into account,
 and obtain another $P^{\rm fun}/P^{\rm 1st}$ versus $L/\pi H$ curve, 
shown by the black line in Figure~\ref{fig_p1p2}.
The difference between these two curves caused by the nonuniform magnetic field is evident, and the $P^{\rm fun}/P^{\rm 1st}$ even exceeds 2 in the black line when $L/\pi H$ is small.

The period ratio exceeding 2 seems consistent with the results of \citet{2008A&A...486.1015V},
who showed that loop expansion can increase the period ratio.
In the current model, 
the loop radius is kept constant in the $y$-direction
but varies in the $\hat{e}_v$ direction perpendicular to the magnetic field lines.
As a result, 
only vertically polarized modes can feel the cross-sectional expansion from the footpoints to the loop apex.
Consequently, 
the increase in radius in the $\hat{e}_{\rm v}$-direction leads to a period ratio of the vertical polarizations greater than 2.

We then proceed to infer the density scale height.
For the horizontally polarized case,
we compute the period ratio between the fundamental mode and the first overtone as $P^{\rm fun}_{\rm h}/P^{\rm 1st}_{\rm h}=1.94$.
This period ratio corresponds to a scale height of approximately 40Mm in our results,
but exceeds 200Mm according to the curve of \citet{2005ApJ...624L..57A},
which is significantly larger than the actual scale height (60–70Mm) of the model.
This indicates that magnetic field inhomogeneity should be included in this inversion scheme.
However, 
if we instead use the period ratio between the vertically polarized fundamental mode and the first overtone, 
namely $P^{\rm fun}_{\rm v}/P^{\rm 1st}_{\rm v}=2.35$,
the scale height can not be derived because this value falls outside the range shown in Figure~\ref{fig_p1p2}.
}

\section{Summary}
\label{sec_summary}

We investigate the response of a gravitationally stratified coronal loop to {different} initial velocity perturbations. 
For the vertically polarized {fundamental mode}, 
the oscillation frequency deviates from the WKB approximation by approximately $18\%$. 
Nevertheless, 
the measured oscillation frequency aligns closely with the local Alfv\'en frequency near the loop apex.  
For the {horizontally polarized fundamental mode}, 
the oscillation frequency is well reproduced by the WKB theory,
with a deviation of only $7\%$. 
Moreover,
the frequency closely matches local Alfv\'en frequency near one quarter of the loop.
These results suggest that the local Alfv\'en frequency at the apex (quarter) of the loop can serve as a more accurate prediction of the frequency of vertically (horizontally) polarized {fundamental} kink oscillations in gravitationally stratified coronal loops.
{For the first overtones,
the oscillation frequencies for both polarizations are close to the WKB approximation,
and their frequencies are closely match the local Alfv\'en frequency close to one quarter of the loop.}

Our findings open the possibility for spatially seismological inference of the coronal magnetic field.
The measured periods of {fundamental} oscillations for different polarizations provide the corresponding local {Alfv\'en frequencies} and,
consequently,
could enable the probing of the magnetic field strength at multiple locations along the loop,
provided that the local density can be obtained in observations.

{
In addition, 
the deviation of the period ratio between the fundamental mode and the first overtone, namely $P^{\rm fun}/P^{\rm 1st}$,
can be used to probe the density scale height in a stratified atmosphere.
By comparison with the results of \cite{2005ApJ...624L..57A},
we find that magnetic field inhomogeneity should be taken into account to obtain a more accurate estimate of the scale height.}

\begin{acknowledgements}
     This work is supported by the National Natural Science Foundation of China (12203030, 12373055, 12273019). 
     M.G. acknowledges the support from the National Natural Science Foundation of China, the QILU Young Scholars Program of Shandong University, the Taishan Scholars Program Special Fund (tsqn202408051) and the Shandong Provincial Natural Science Foundation for Excellent Young Scientists Program, Overseas (2025HWYQ-019).
\end{acknowledgements}

%
\bibliographystyle{aa} 
\bibliography{ref} 

\begin{thebibliography}{43}
\expandafter\ifx\csname natexlab\endcsname\relax\def\natexlab#1{#1}\fi

\bibitem[{{Andries} {et~al.}(2005{\natexlab{a}}){Andries}, {Arregui}, \&
  {Goossens}}]{2005ApJ...624L..57A}
{Andries}, J., {Arregui}, I., \& {Goossens}, M. 2005{\natexlab{a}}, \apjl, 624,
  L57

\bibitem[{{Andries} {et~al.}(2005{\natexlab{b}}){Andries}, {Goossens},
  {Hollweg}, {Arregui}, \& {Van Doorsselaere}}]{2005A&A...430.1109A}
{Andries}, J., {Goossens}, M., {Hollweg}, J.~V., {Arregui}, I., \& {Van
  Doorsselaere}, T. 2005{\natexlab{b}}, \aap, 430, 1109

\bibitem[{{Brady} \& {Arber}(2005)}]{2005A&A...438..733B}
{Brady}, C.~S. \& {Arber}, T.~D. 2005, \aap, 438, 733

\bibitem[{{Dymova} \& {Ruderman}(2006)}]{2006A&A...457.1059D}
{Dymova}, M.~V. \& {Ruderman}, M.~S. 2006, \aap, 457, 1059

\bibitem[{{Edwin} \& {Roberts}(1983)}]{1983SoPh...88..179E}
{Edwin}, P.~M. \& {Roberts}, B. 1983, \solphys, 88, 179

\bibitem[{{Erd{\'e}lyi} \& {Verth}(2007)}]{2007A&A...462..743E}
{Erd{\'e}lyi}, R. \& {Verth}, G. 2007, \aap, 462, 743

\bibitem[{{Gao} {et~al.}(2025){Gao}, {Tian}, {Van Doorsselaere}, {Yang}, {Guo},
  \& {Karampelas}}]{2025RAA....25a5010G}
{Gao}, Y., {Tian}, H., {Van Doorsselaere}, T., {et~al.} 2025, Research in
  Astronomy and Astrophysics, 25, 015010

\bibitem[{{Gao} {et~al.}(2024){Gao}, {Van Doorsselaere}, {Tian}, {Guo}, \&
  {Karampelas}}]{2024A&A...689A.195G}
{Gao}, Y., {Van Doorsselaere}, T., {Tian}, H., {Guo}, M., \& {Karampelas}, K.
  2024, \aap, 689, A195

\bibitem[{{Goossens} {et~al.}(2002){Goossens}, {Andries}, \&
  {Aschwanden}}]{2002A&A...394L..39G}
{Goossens}, M., {Andries}, J., \& {Aschwanden}, M.~J. 2002, \aap, 394, L39

\bibitem[{{Goossens} {et~al.}(1992){Goossens}, {Hollweg}, \&
  {Sakurai}}]{1992SoPh..138..233G}
{Goossens}, M., {Hollweg}, J.~V., \& {Sakurai}, T. 1992, \solphys, 138, 233

\bibitem[{{Guo} {et~al.}(2023){Guo}, {Duckenfield}, {Van Doorsselaere},
  {Karampelas}, {Pelouze}, \& {Gao}}]{2023ApJ...949L...1G}
{Guo}, M., {Duckenfield}, T., {Van Doorsselaere}, T., {et~al.} 2023, \apjl,
  949, L1

\bibitem[{{Guo} {et~al.}(2020){Guo}, {Li}, \& {Van
  Doorsselaere}}]{2020ApJ...904..116G}
{Guo}, M., {Li}, B., \& {Van Doorsselaere}, T. 2020, \apj, 904, 116

\bibitem[{{Guo} {et~al.}(2024){Guo}, {Van Doorsselaere}, {Li}, \&
  {Goossens}}]{2024A&A...687A..30G}
{Guo}, M., {Van Doorsselaere}, T., {Li}, B., \& {Goossens}, M. 2024, \aap, 687,
  A30

\bibitem[{{Karampelas} \& {Van Doorsselaere}(2021)}]{2021ApJ...908L...7K}
{Karampelas}, K. \& {Van Doorsselaere}, T. 2021, \apjl, 908, L7

\bibitem[{{Karampelas} {et~al.}(2019){Karampelas}, {Van Doorsselaere}, \&
  {Guo}}]{2019A&A...623A..53K}
{Karampelas}, K., {Van Doorsselaere}, T., \& {Guo}, M. 2019, \aap, 623, A53

\bibitem[{{Lopin} \& {Nagorny}(2024)}]{2024MNRAS.527.5741L}
{Lopin}, I. \& {Nagorny}, I. 2024, \mnras, 527, 5741

\bibitem[{{Lopin} {et~al.}(2014){Lopin}, {Nagorny}, \&
  {Nippolainen}}]{2014SoPh..289.3033L}
{Lopin}, I.~P., {Nagorny}, I.~G., \& {Nippolainen}, E. 2014, \solphys, 289,
  3033

\bibitem[{{Magyar} \& {Nakariakov}(2020)}]{2020ApJ...894L..23M}
{Magyar}, N. \& {Nakariakov}, V.~M. 2020, \apjl, 894, L23

\bibitem[{{Mignone} {et~al.}(2007){Mignone}, {Bodo}, {Massaglia}, {Matsakos},
  {Tesileanu}, {Zanni}, \& {Ferrari}}]{2007ApJS..170..228M}
{Mignone}, A., {Bodo}, G., {Massaglia}, S., {et~al.} 2007, \apjs, 170, 228

\bibitem[{{Nakariakov} \& {Kolotkov}(2020)}]{2020ARA&A..58..441N}
{Nakariakov}, V.~M. \& {Kolotkov}, D.~Y. 2020, \araa, 58, 441

\bibitem[{{Nakariakov} \& {Ofman}(2001)}]{2001A&A...372L..53N}
{Nakariakov}, V.~M. \& {Ofman}, L. 2001, \aap, 372, L53

\bibitem[{{Pelouze} {et~al.}(2023){Pelouze}, {Van Doorsselaere}, {Karampelas},
  {Riedl}, \& {Duckenfield}}]{2023A&A...672A.105P}
{Pelouze}, G., {Van Doorsselaere}, T., {Karampelas}, K., {Riedl}, J.~M., \&
  {Duckenfield}, T. 2023, \aap, 672, A105

\bibitem[{{Reale}(2014)}]{2014LRSP...11....4R}
{Reale}, F. 2014, Living Reviews in Solar Physics, 11, 4

\bibitem[{{Ruderman}(2009)}]{2009A&A...506..885R}
{Ruderman}, M.~S. 2009, \aap, 506, 885

\bibitem[{{Selwa} {et~al.}(2005){Selwa}, {Murawski}, {Solanki}, {Wang}, \&
  {T{\'o}th}}]{2005A&A...440..385S}
{Selwa}, M., {Murawski}, K., {Solanki}, S.~K., {Wang}, T.~J., \& {T{\'o}th}, G.
  2005, \aap, 440, 385

\bibitem[{{Smith} {et~al.}(1997){Smith}, {Roberts}, \&
  {Oliver}}]{1997A&A...317..752S}
{Smith}, J.~M., {Roberts}, B., \& {Oliver}, R. 1997, \aap, 317, 752

\bibitem[{{Spruit}(1981)}]{1981A&A....98..155S}
{Spruit}, H.~C. 1981, \aap, 98, 155

\bibitem[{{Terradas} {et~al.}(2006){Terradas}, {Oliver}, \&
  {Ballester}}]{2006ApJ...650L..91T}
{Terradas}, J., {Oliver}, R., \& {Ballester}, J.~L. 2006, \apjl, 650, L91

\bibitem[{{Van Doorsselaere} {et~al.}(2004){Van Doorsselaere}, {Debosscher},
  {Andries}, \& {Poedts}}]{2004A&A...424.1065V}
{Van Doorsselaere}, T., {Debosscher}, A., {Andries}, J., \& {Poedts}, S. 2004,
  \aap, 424, 1065

\bibitem[{{Van Doorsselaere} {et~al.}(2007){Van Doorsselaere}, {Nakariakov}, \&
  {Verwichte}}]{2007A&A...473..959V}
{Van Doorsselaere}, T., {Nakariakov}, V.~M., \& {Verwichte}, E. 2007, \aap,
  473, 959

\bibitem[{{Van Doorsselaere} {et~al.}(2009){Van Doorsselaere}, {Verwichte}, \&
  {Terradas}}]{2009SSRv..149..299V}
{Van Doorsselaere}, T., {Verwichte}, E., \& {Terradas}, J. 2009, \ssr, 149, 299

\bibitem[{{Verth} \& {Erd{\'e}lyi}(2008)}]{2008A&A...486.1015V}
{Verth}, G. \& {Erd{\'e}lyi}, R. 2008, \aap, 486, 1015

\bibitem[{{Verwichte} {et~al.}(2006{\natexlab{a}}){Verwichte}, {Foullon}, \&
  {Nakariakov}}]{2006A&A...446.1139V}
{Verwichte}, E., {Foullon}, C., \& {Nakariakov}, V.~M. 2006{\natexlab{a}},
  \aap, 446, 1139

\bibitem[{{Verwichte} {et~al.}(2006{\natexlab{b}}){Verwichte}, {Foullon}, \&
  {Nakariakov}}]{2006A&A...449..769V}
{Verwichte}, E., {Foullon}, C., \& {Nakariakov}, V.~M. 2006{\natexlab{b}},
  \aap, 449, 769

\bibitem[{{Verwichte} {et~al.}(2006{\natexlab{c}}){Verwichte}, {Foullon}, \&
  {Nakariakov}}]{2006A&A...452..615V}
{Verwichte}, E., {Foullon}, C., \& {Nakariakov}, V.~M. 2006{\natexlab{c}},
  \aap, 452, 615

\bibitem[{{Wang} \& {Solanki}(2004)}]{2004A&A...421L..33W}
{Wang}, T.~J. \& {Solanki}, S.~K. 2004, \aap, 421, L33

\bibitem[{{Wang} {et~al.}(2008){Wang}, {Solanki}, \&
  {Selwa}}]{2008A&A...489.1307W}
{Wang}, T.~J., {Solanki}, S.~K., \& {Selwa}, M. 2008, \aap, 489, 1307

\bibitem[{{Yang} {et~al.}(2020{\natexlab{a}}){Yang}, {Bethge}, {Tian},
  {Tomczyk}, {Morton}, {Del Zanna}, {McIntosh}, {Karak}, {Gibson}, {Samanta},
  {He}, {Chen}, \& {Wang}}]{2020Sci...369..694Y}
{Yang}, Z., {Bethge}, C., {Tian}, H., {et~al.} 2020{\natexlab{a}}, Science,
  369, 694

\bibitem[{{Yang} {et~al.}(2024){Yang}, {Tian}, {Tomczyk}, {Liu}, {Gibson},
  {Morton}, \& {Downs}}]{2024Sci...386...76Y}
{Yang}, Z., {Tian}, H., {Tomczyk}, S., {et~al.} 2024, Science, 386, 76

\bibitem[{{Yang} {et~al.}(2020{\natexlab{b}}){Yang}, {Tian}, {Tomczyk},
  {Morton}, {Bai}, {Samanta}, \& {Chen}}]{2020ScChE..63.2357Y}
{Yang}, Z., {Tian}, H., {Tomczyk}, S., {et~al.} 2020{\natexlab{b}}, Science in
  China E: Technological Sciences, 63, 2357

\bibitem[{{Zajtsev} \& {Stepanov}(1975)}]{1975IGAFS..37....3Z}
{Zajtsev}, V.~V. \& {Stepanov}, A.~V. 1975, Issledovaniia Geomagnetizmu
  Aeronomii i Fizike Solntsa, 37, 3

\bibitem[{{Zhong} {et~al.}(2023){Zhong}, {Nakariakov}, {Kolotkov}, {Chitta},
  {Antolin}, {Verbeeck}, \& {Berghmans}}]{2023NatCo..14.5298Z}
{Zhong}, S., {Nakariakov}, V.~M., {Kolotkov}, D.~Y., {et~al.} 2023, Nature
  Communications, 14, 5298

\bibitem[{{Zimovets} \& {Nakariakov}(2015)}]{2015A&A...577A...4Z}
{Zimovets}, I.~V. \& {Nakariakov}, V.~M. 2015, \aap, 577, A4

\end{thebibliography}
%

\end{document}